\documentclass[
    ,final            
  ]
  {aipproc}

\layoutstyle{8x11single}

\begin{document}

\title{GRB 081029: Understanding Multiple Afterglow Components}


\author{S. T. Holland}{
  address={Centre for Research and Exploration in Space Science and Technology,
           NASA/GSFC,
           Greenbelt, MD 20771, USA},
  altaddress={Universities Space Research Association,
              10211 Winicopin Circle,
              Columbia, MD 21044, USA},
  altaddress={Code 660.1, NASA/GSFC,
              Greenbelt, MD 20771, USA}
}

\author{M. De Pasquale}{
  address={Mullard Space Science Laboratory, University College London,
           Holmbury, St Mary, Dorking, Surrey, RH5~6NT, UK}
}

\author{J. Mao}{
  address={INAF-Osservatorio Astronomico di Brera, Via Emilio Bianchi 46,
           I--23807 Merate (LC), Italy}
}

\author{T. Sakamoto}{
  address={Centre for Research and Exploration in Space Science and Technology,
           NASA/GSFC,
           Greenbelt, MD 20771, USA},
  altaddress={Joint Centre for Astrophysics,
              University of Maryland, Baltimore County,
              1000 Hilltop Circle, Baltimore, MD 21250, USA},
  altaddress={Code 660.1, NASA/GSFC,
              Greenbelt, MD 20771, USA}
}

\author{P. Schady}{
  address={Max-Planck Institut f{\"u}r Extraterrestrische Physik,
           Giessenbachstra{\ss}e, 85748 Garching, Germany},
  altaddress={Mullard Space Science Laboratory, University College London,
              Holmbury, St Mary, Dorking, Surrey, RH5~6NT, UK}
}

\author{S. Covino}{
  address={INAF-Osservatorio Astronomico di Brera, Via Emilio Bianchi 46,
           I--23807 Merate (LC), Italy}
}

\author{P. D'Avanzo}{
  address={INAF-Osservatorio Astronomico di Brera, Via Emilio Bianchi 46,
           I--23807 Merate (LC), Italy}
}

\author{A. Antonelli}{
  address={INAF-Osservatorio Astronomico di Roma, Via de Frascati 33,
           I--00040 Monteporzio Catone (Roma), Italy}
}

\author{V. D'Elia}{
  address={INAF-Osservatorio Astronomico di Roma, Via de Frascati 33,
           I--00040 Monteporzio Catone (Roma), Italy}
}

\author{G. Chincarini}{
  address={INAF-Osservatorio Astronomico di Brera, Via Emilio Bianchi 46,
           I--23807 Merate (LC), Italy}
}

\author{F. Fiore}{
  address={INAF-Osservatorio Astronomico di Roma, Via de Frascati 33,
           I--00040 Monteporzio Catone (Roma), Italy}
}

\author{S. B. Pandey}{
  address={Randall Laboratory of Physics, University of Michigan,
           450 Church St, Ann Arbor, MI 48109--1040, USA} 
}

\begin{abstract}
  We present an analysis of the unusual optical light curve of the
  gamma-ray burst GRB~081029, which occurred at a redshift of $z =
  3.8479$.  We combine $X$-ray and optical observations from the {\sl
    Swift\/} $X$-Ray Telescope and the {\sl Swift\/} UltraViolet
  Optical Telescope with optical and infrared data obtained using the
  REM and ROTSE telescopes to construct a detailed data set extending
  from 86~s to $\sim$100\,000~s after the BAT trigger.  Our data also
  cover a wide energy range, from 10~keV to 0.77~eV (1.24~{\AA} to
  16\,000~{\AA}).  The $X$-ray afterglow shows a shallow initial decay
  followed by a rapid decay starting at about 18\,000~s.  The optical
  and infrared afterglow, however, shows an uncharacteristic rise at
  about 5000~s that does not correspond to any feature in the $X$-ray
  light curve.  Our data are not consistent with synchrotron radiation
  from a single-component jet interacting with an external medium.  We
  do, however, find that the observed light curve can be explained
  using multi-component model for the jet.
\end{abstract}

\maketitle


\section{Introduction}

There is growing evidence that the classical picture of a single
uniform jet cannot explain the spectral energy distributions and light
curves of some gamma-ray burst (GRB) afterglows.  For example, the
unusually bright optical afterglow of the ``naked-eye'' burst
GRB~080319B was best explained using a two-component jet
\citep{RKS2008} while GRB~030329 \citep{BKP2003} appears to require a
narrow, ultra-relativistic inner jet and a wide, mildly relativistic
outer jet to explain its light curves.  This is in agreement with
results from magneto-hydrodynamic modelling that show complex structure
in GRB jets \citep[e.g.,][]{TNM2010}.  GRB afterglows appear to be
more complex than originally thought.

An example of a GRB afterglow that appears to require a
multi-component jet is GRB~081029.  This burst was detected by {\sl
  Swift\/}/BAT at 01:43:56 UT on 2008 Oct 29 \citep{SBB2008}.
ROTSE-IIIc identified the optical afterglow at 86~s \citep{R2008}, and
the REM telescope started observing the optical afterglow at 154~s
\citep{CAM2008}, so there is a well-sampled $R$-band light curve
starting less than 90~s after the BAT trigger.  Due to an observing
constraing {\sl Swift\/} was unable to slew to this burst as soon as
it was detected.  XRT and UVOT observations began about 45 minute
after the BAT trigger and continued for approximately 10 days.

The VLT/UVES and Gemini-South measured a redshift of $z = 3.8479$
\citep{DCD2008,CFC2008}, which corresponds to a look back time of
11.9~Gyr.  The Gemini-South GMOS spectrum shows evidence for a damped
Lyman-alpha system as well as several metal absorption features in the
host galaxy of GRB~081029.

\section{Observations}

\subsection{BAT: Prompt Emission}

\begin{figure}[h]
  \includegraphics[height=0.3\textheight,angle=-90]{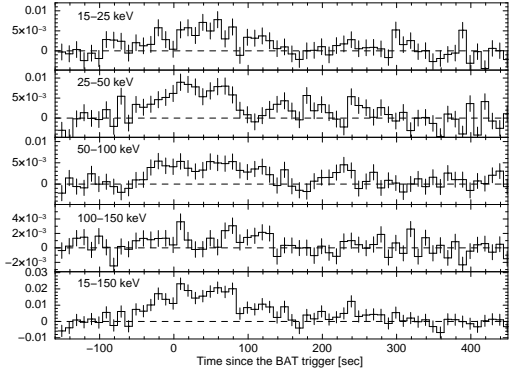}
  \caption{The BAT energy resolved light curves of GRB~081029 with
    10~s binning.}\label{FIGURE:bat}
\end{figure}

The {\sl Swift\/}/BAT discovered and observed GRB~081029.  The burst
duration was $T_{90} = 280 \pm 50$~s, the peak flux was $(2.8 \pm 1.3)
\times 10^{-8}$~erg~cm$^{-2}$~s$^{-1}$, and the spectrum was best fit
by a simple power law with a photon index of $\Gamma = 1.5 \pm 0.2$.
The BAT light curve was somewhat smoother and weaker than a typical
BAT-detected GRB.  Figure~\ref{FIGURE:bat} shows the BAT light cuvrve
for the prompe emission from GRB~081029.

\subsection{XRT: X-Ray Light Curve and Spectrum}

\begin{figure}[h]
  \includegraphics[height=.3\textheight]{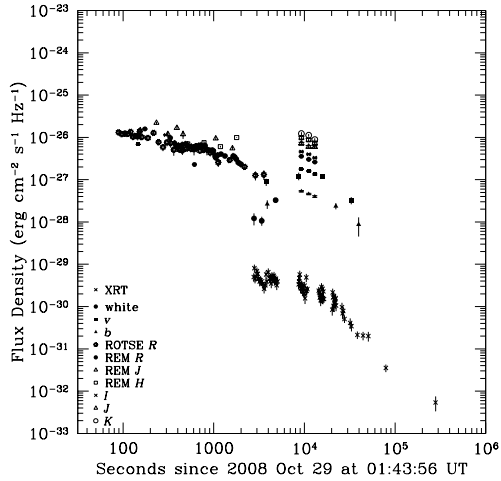}
  \caption{The $X$-ray, optical, and infrared data presented as flux
    densities.}\label{FIGURE:flux_densities}
\end{figure}

The {\sl Swift\/}/XRT observed GRB~081029 from 41.4~minutes to
approximately 10~days after the BAT trigger.  The $X$-ray light curve
(see Figure~\ref{FIGURE:flux_densities}) is well fit by a broken power
law with indices ($f_{\nu} \propto t^{-\alpha}$) of $\alpha_1 = 0.56
\pm 0.03$ until $18\,230 \pm 346$~s, and $\alpha_2 = 2.56 \pm 0.09$
after that.  There is some evidence for flaring between approximately
2500~s and 5000~s.  The time scales of these flares are consistent
with $\Delta t / t < 1$.  The $X$-ray data are not unusual, and are
consistent with the canonical $X$-ray light curve for GRB afterglows
described by \citep{NKG2006} and \citep{ZFD2006}.

The {\sl Swift\/}/XRT spectrum can be fit by a single power law
($f_{\nu} \propto \nu^{-\beta}$) with an index of $\beta_X = 0.98 \pm
0.08$.  There is no evidence for any evolution in the power law index
at $X$-ray energies.  The estimated Galactic column density in the
direction of the burst is $N_H = 2.8 \times 10^{-20}$~cm$^{-2}$, and
the absorption in the host is $N_H = 4.9 \times 10^{-21}$~cm$^{-2}$.

\subsection{Optical and Infrared Observations}

The {\sl Swift\/}UVOT began observing the afterglow of GRB~081029 at
2689~s after the BAT trigger.  The afterglow was detected in the UVOT
$v$, $b$, and white bands, consistent with the reported redshift of $z
= 3.8479$.  Ground-based data was obtained using REM and ROTSE\@.
ROTSE began observations 86~s after the burst in the $R$ band.  REM
began observing GRB~081029 156~s after the BAT trigger in the $R$,
$J$, and $H$ bands.  The resulting light curves are complex, in stark
contrast to the simple $X$-ray light curve.  The combined optical and
infrared observations are shown in Figure~\ref{FIGURE:flux_densities}
along with the {\sl Swift\/}/XRT light curve.  The optical and
infrared data show a jump in the flux density of approximately a
factor of ten at approximately 5000~s.  There is no corresponding
increase in the $X$-ray flux density at that time.

\section{Discussion}

The $X$-ray light curve is consistent with energy injection from
ongoing central engine activity until about 15\,000~s followed by a
jet break at 18\,230~s.  However, this scenario cannot explain the
jump in the flux seen at optical and infrared wavelengths at about
5000~s.  Therefore, we do not think that a change in the energy
injection is capable of explaining the light curves for this
afterglow.  Similarly, the jump cannot be modelled by invoking the
passage of the synchrotron peak frequency through the optical regime,
or as the rise of the forward shock due to interaction with the
circumburst medium.  We also examined the possibility that the jump is
due to density structure in the surrounding environment, but this is
unable to reproduce the speed or the magnitude of the increase in
luminosity.

\begin{figure}
  \includegraphics[height=0.4\textheight]{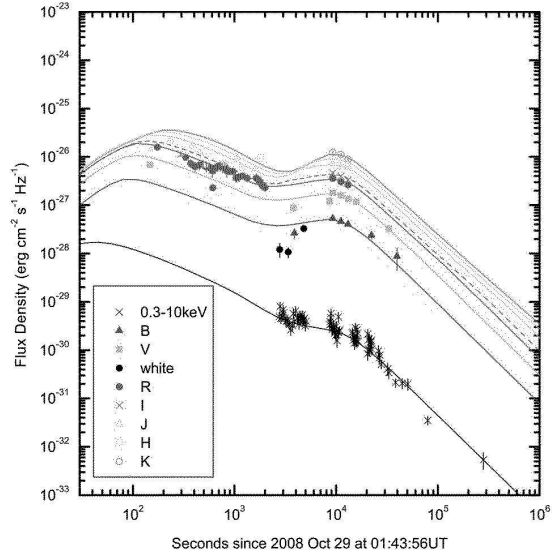}
  \caption{A two-component jet model provides a reasonable fit to the
    optical, infrared, and $X$-ray light curves of the afterglow of
    GRB~081029.}\label{FIGURE:model}
\end{figure}

In general we find that a one-component jet cannot explain the
observed light curves and spectral energy distribution of the $X$-ray,
optical, and infrared afterglows of GRB~081029.  However, a
two-component jet model, similar to what is seen in some other GRB
afterglows, does provide a reasonable fit to the data Our
two-component jet model is shown in Figure~\ref{FIGURE:model}, and the
parameters of each jet are listed in Table~\ref{TABLE:model}.  The
half-opening angle of the jet is denoted by $\theta_j$, $\Gamma_0$ is
the Lorentz factor, $E_{K,\mathrm{iso}}$ is the isotropic equivalent
kinetic energy in the jet, $p$ is the electron index, $\epsilon_e$ and
$\epsilon_B$ are the fractions of the energy in electrons and magnetic
fields respectively, $n$ is the density of the circumburst medium, and
$z$ is the redshift.

\begin{table}
\begin{tabular}{ccc}
\hline
    \tablehead{1}{c}{b}{Parameter}
  & \tablehead{1}{c}{b}{Narrow Jet}
  & \tablehead{1}{c}{b}{Wide Jet} \\
\hline
$\theta_j$ (rad)         &  0.01               & 0.02 \\
$\Gamma_0$               &  500                & 60 \\
$E_{K,\mathrm{iso}}$ (erg) & $2.5 \times 10^{54}$ & $2.0 \times 10^{54}$ \\
$p$                      & 2.2                 & 2.2 \\
$\epsilon_e$             & 0.02                & 1/3 \\
$\epsilon_B$             & 0.0002              & 0.0002 \\
$n$ (cm$^{-3}$)           & 10                  & 10 \\
$z$                      & 3.8479              & 3.8479 \\
\hline
\end{tabular}
\caption{Model parameters for the best-fitting two-component jet model for
  GRB~081029.}
\label{TABLE:model}
\end{table}

The narrow, inner jet has a half-opening angle of $\theta_{j,n} =
0.01$~rad and a Lorentz factor of 500.  This component gives rise to
the $X$-ray flux and the pre-jump optical flux.  The wider, outer jet
has $\theta_{j,w} = 0.02$~rad and a Lorentz factor of 60.  This
component dominates the afterglow after about 10\,000~s.  The total
electromagnetic energy in the afterglow is approximately equally
divided between the two jets.

\section{Conclusions}

GRB~081029 was a long--soft GRB with a redshift of $z = 3.8479$.  It
had a smooth gamma-ray light curve and did not appear to have any
unusual gamma-ray properties.  Neither the gamma-ray nor the $X$-ray
properties of this burst showed any sign of strange behaviour.  The
optical and infrared light curves, on the other hand, were not typical
of GRB afterglows.  There is a brightening in the optical and infrared
light curves at about 5000~s that cannot be explained using a
single-component jet model.  However, we find that a two-component jet
model fits the data reasonably well.

We conclude that the afterglow of GRB~081029 was probably powered by a
two-component jet with the energy split approximately equally between
a narrow ($\theta_{j,n} = 0.01$~rad) inner jet and a wider
($\theta_{j,w} = 0.02$~rad) outer jet.  The inner jet has a Lorentz
factor of $\Gamma_n = 500$ while the outer jet has $\Gamma_w = 60$.
This result provides evidence that some (and perhaps all) GRB jets
have complex internal structure.


\begin{theacknowledgments}
  We acknowledge the use of public data from the {\sl Swift\/} Data
  Archive.  This work is based in part on observations taken with the
  ROTSE-IIIc telescope in Namibia, the REM telescope at la Silla
  Observatory, and with ESO Telescopes at the Paranal Observatories.
\end{theacknowledgments}

\end{document}